\documentclass[a4paper,11pt]{article}
\usepackage{pos}
\usepackage{pgfplots}

\DeclareMathOperator{\tr}{tr}
\DeclareMathOperator{\Tr}{Tr}
\DeclareMathOperator{\Link}{Link}

\title{Higher-group symmetry in lattice gauge theories with restricted topological sectors}

\author[a]{Motokazu Abe}
\author[b]{Naoto Kan}
\author*[b]{Okuto Morikawa}
\author[b]{Yuta Nagoya}
\author[a]{Soma Onoda}
\author[b]{Hiroki Wada}

\affiliation[a]{Department of Physics, Kyushu University,
744 Motooka, Nishi-ku, Fukuoka 819-0395, Japan}

\affiliation[b]{Department of Physics, Osaka University,
Toyonaka, Osaka 560-0043, Japan}

\emailAdd{abe.motokazu@phys.kyushu-u.ac.jp}
\emailAdd{naotokan000@gmail.com}
\emailAdd{o-morikawa@het.phys.sci.osaka-u.ac.jp}
\emailAdd{y\_nagoya@het.phys.sci.osaka-u.ac.jp}
\emailAdd{onoda.soma.270@s.kyushu-u.ac.jp}
\emailAdd{hwada@het.phys.sci.osaka-u.ac.jp}

\abstract{
In this paper, we give a brief overview of generalized symmetries
from the point of view of the lattice regularization as a fully regularized framework.
At first, we illustrate the generalization of 't~Hooft anomaly matching
for higher-form symmetries.
Furthermore the main interest goes to the higher-group symmetry.
In particular, we find that the so-called $4$-group appears
in the lattice Yang--Mills theory under modification of instanton sum.
}

\FullConference{The 40th International Symposium on Lattice Field Theory (Lattice 2023)\\
July 31st - August 4th, 2023\\
Fermi National Accelerator Laboratory\\}

\begin{document}
\maketitle

\section{Introduction and summary}
Every physicist has cultivated intuition about symmetry.
Having established a paradigm in triumph,
we understand more aspects of physics,
such as conservation laws, phase transitions, and fundamental forces in the nature.
Anomalous or spontaneous breaking of symmetries also plays an essential role
to clarify theoretical (and experimental) consistency,
and/or to classify the phase structure since the Landau theory.
The 't~Hooft anomaly~\cite{tHooft:1979rat},
a renormalization-group-invariant quantum anomaly
arising from gauging a global symmetry,
tells us quite nontrivial restrictions on the low-energy dynamics and vacuum structure.
This idea of anomaly matching is applicable to the recent generalization
of symmetry~\cite{Gaiotto:2014kfa,Gaiotto:2017yup},
and thus can improve our insight
about nonperturbative phenomena.\footnote{%
For studies of gauge theories, see Refs.~\cite{Sulejmanpasic:2020zfs,Morikawa:2022liz}
and references cited therein.}

First of all, in Sect.~\ref{sec:review},
let us review the basic concept of generalized symmetries:
higher-form symmetry which leads us to
a generalized 't~Hooft anomaly in the $SU(N)$ Yang--Mills theory with the $\theta$ term.
To aim at transparent understanding of it,
we make remarks on topology within a fully regularized framework
given in Refs.~\cite{Abe:2022nfq,Abe:2023ncy};
its description enjoys the topological structure
\emph{even on a lattice}~\cite{Luscher:1981zq}.
In this paper, we focus on higher-group symmetry (see Sect.~\ref{sec:higher-group}).
For instance, on background Abelian gauge fields $A_\mu$ and~$B_{\mu\nu}$,
the gauge transformation acts as $A\mapsto A+d\omega$, $B\mapsto B+d\lambda+\omega d A$,
and then such a mixture of symmetries is called the $2$-group;
as we know, this is similar to the Green--Schwarz mechanism.\footnote{%
In the case of the Green--Schwarz mechanism, both gauge fields are dynamical rather than background.}
From our lattice viewpoint of generalized symmetries,
as in Refs.~\cite{Kan:2023yhz,Abe:2023ubg},
we can construct a $4$-group structure in the \emph{lattice} $SU(N)$ Yang--Mills theory
with restricted topological sectors~\cite{Tanizaki:2019rbk}.

Other kinds of generalized symmetries are still developing.
We hope to apply our approach to recent developments as
non-invertible symmetry, subsystem symmetry and so on.

\section{Understanding of generalized symmetries within a fully regularized framework}
\label{sec:review}
\subsection{Higher-form symmetry and 't~Hooft anomaly}
The basic notion of generalized symmetries is as follows:
\begin{enumerate}
 \item The concept of symmetry is regarded as a topological defect.
       As depicted in Fig.~\ref{fig:higher-form},
       for an ordinary ($0$-form) symmetry we have the symmetry defect operator
       $U_\alpha(\Sigma)$ on the codimension-$1$ space~$\Sigma$ with fixed time,
       \begin{align}
	Q&\equiv\int_{\Sigma_{D-1}} j_0 d x_1\wedge\dots\wedge d x_{D-1}, &
	U_\alpha(\Sigma_{D-1})&\equiv e^{i\alpha Q},
       \end{align}
       while a generic symmetry defect on the codimension-$(p+1)$ surface
       is given by
       \begin{align}
	Q&\equiv \int_{\Sigma_{D-p-1}} \star j^{(p+1)}, &
	U_\alpha(\Sigma_{D-p-1}) &\equiv e^{i\alpha Q} . \label{eq:sym_op}
       \end{align}
\begin{figure}[t]
 \begin{center}
 \begin{tikzpicture}
  \draw[->] (0,-0.2) -- (0,2) node[right] {$t=x_0$};
  \fill (0,1) circle(2pt);
  \draw[very thick,blue] (1,0,-2) .. controls (0.8,-0.5,-2) and (1.2,-0.5,-2) .. (1,-1,-2) node[below] {$1$D};
  \fill[opacity=0.8,black!50] (1,0,0) -- (3,0,-1) -- (0.5,0,-5) -- (-1,0,-4);
  \fill[blue] (1,0,-2) circle(2pt);
  \draw[very thick,blue] (1,0,-2) .. controls (1.5,1,-2) and (0.5,1,-2) .. (1,1.5,-2);
  \node[above right,black] at (2.5,0,-1.5) {$\Sigma_{D-1}$};
 \end{tikzpicture}
  \hspace{1em}
 \begin{tikzpicture}
  \draw[very thick,black] (0,1,-2) .. controls (-0.5,1,-1.5) and (-1,1,-2.5) .. (-1.2,1,-2);
  \fill[opacity=0.8,blue!50] (0,0,0) -- (0,0,-4) -- (0,2,-4) -- (0,2.5,0);
  \node[above right,blue] at (0.3,0) {$(p+1)$D};
  \draw[very thick,black] (0,1,-2) .. controls (0.5,1,-2.5) and (1,1,-1.5) .. (1.5,1,-2) node[above] {$\Sigma_{D-p-1}$};
  \fill[black] (0,1,-2) circle(2pt);
 \end{tikzpicture}
  \caption{Generalization of symmetry as a topological defect.
  Symmetry operator (or charge) is defined on the codimension-$(p+1)$ defect~$\Sigma$,
  as $U_\alpha(\Sigma)$ in eq.~\eqref{eq:sym_op}.}
  \label{fig:higher-form}
 \end{center}
\end{figure}
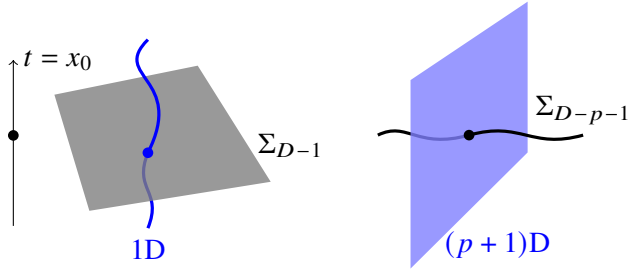
 \item The above $p$-form symmetry with $p\geq1$ is Abelian.
       In particular we are interested in discrete global symmetries.
       For instance, the $SU(N)$ Yang--Mills theory possesses
       the $\mathbb{Z}_N$ $1$-form ($\mathbb{Z}_N^{[1]}$) center symmetry;
       Fig.~\ref{fig:center_lat} shows its defect from the lattice viewpoint.
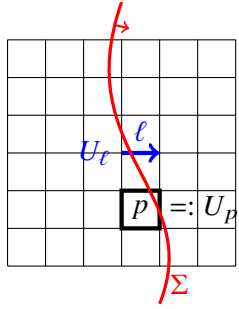
\begin{figure}[t]
 \begin{center}
 \begin{tikzpicture}
  \draw (0,0) grid[step=0.5] (3,3);
  \draw[ultra thick,blue,->] (1.5,1.5) -- (2,1.5) node [midway,above] {$\ell$};
  \node[left,blue] at (1.5,1.5) {$U_\ell$};
  \draw[ultra thick] (1.5,0.5) rectangle (2,1) node[midway] {$p$};
  \node[right] at (2,0.75) {$=:U_p$};
  \draw[very thick,red] (2,-0.5) .. controls (2.6,1) and (0.8,1.5) .. (1.5,3.5);
  \draw[thick,red,->] (1.4,3.2) -- (1.6,3.15);
  \node[red,right] at (2,-0.25) {$\Sigma$};
 \end{tikzpicture}
  \caption{Center of gauge group as an example of $1$-form symmetry.
  The lattice $\Lambda$ divides a four-dimensional torus into hypercubes, and
  $\Sigma$ is a codimension-$1$ oriented surface such that $\Sigma\cup\Lambda=\emptyset$.
  One may regard $\Sigma$ as a network on the dual lattice.
  Under a center transformation on~$\Sigma$, any plaquette is invariant,
  whereas $\partial\Sigma$ can provide the codimension-$2$ symmetry defect.}
  \label{fig:center_lat}
 \end{center}
\end{figure}
 \item A charged object $W(C)$ is defined as a ``loop'' operator,
       and transforms under action of the symmetry operator by
       \begin{align}
	\left\langle W(C)\right\rangle \mapsto\left\langle U(\Sigma)W(C)\right\rangle
	= \left\langle e^{i\alpha\Link(\Sigma,C)} W(C)\right\rangle,
       \end{align}
       where $\Link(\Sigma,C)$ denotes the intersection number
       between the surfaces $\Sigma$ and $C$.
       The center transformation in Fig.~\ref{fig:center_lat}
       is then given by~$U_\ell\mapsto e^{\frac{2\pi i}{N}k\Link(\Sigma,\ell)}U_\ell$
       with any bond $\ell$ and an integer $k$.
       Note that any plaquette (or lattice action) is invariant; $U_p\mapsto U_p$.
 \item To gauge higher-form global symmetries, for example, let us construct
       the $\mathbb{Z}_N^{[1]}$ gauge symmetry from the $SU(N)$ gauge theory.
       The lattice action,
       \begin{align}
	S[U_\ell,B_p]
	= \sum_p\beta\left[\tr\left(1-e^{-\frac{2\pi i}{N}B_p}U_p\right)+\text{c.c.}\right],
	\label{eq:wilson_action}
       \end{align}
       where $B_p$ is a $2$-form gauge field associated
       with~$\mathbb{Z}_N^{[1]}$,\footnote{%
       Some people are familiar with the 't~Hooft twisted boundary condition.
       For $\forall\ell=(n,\mu)$, gauge functions obey
       \begin{align}
	U_{n+L\Hat\nu,\mu} &= g_{n,\nu}^{-1}U_{n,\mu}g_{n+\Hat\mu,\nu}, &
	g_{n+L\Hat\nu,\mu}^{-1}g_{n,\nu}^{-1}g_{n,\mu}g_{n+L\Hat\mu,\nu}
	= e^{\frac{2\pi i}{N}z_{\mu\nu}}\in\mathbb{Z}_N
	\label{eq:twist_bc}
       \end{align}
       where the 't~Hooft flux $z_{\mu\nu}$ is identical to $\sum B_p \bmod N$.
       The topological charge has
       a fractional shift~$\sim\frac{1}{N}$
       due to~$z_{\mu\nu}$~\cite{vanBaal:1982ag}.
       This expression looks to be written by using the global data,
       but it can be written in terms of local operations to~$B_p$
       from a modern perspective as we will see.}
       is invariant under the gauge transformation
       \begin{align}
	U_\ell&\mapsto e^{\frac{2\pi i}{N}\lambda_\ell}U_\ell, &
	B_p&\mapsto B_p + (d\lambda)_p \bmod N.
	\label{eq:1-form_gauge_transf}
       \end{align}
\end{enumerate}

It is convenient to consider the above idea from the lattice viewpoint.
The strategy through the use of 't~Hooft anomaly matching is, however, closely related
to topology of gauge fields and based on cohomological operations.
From now, one would have a tendency to refrain the lattice regularized framework.
To do this, as a formal standpoint in the continuum theory,
\begin{enumerate}
 \item The ``$\mathbb{Z}_N$ $2$-form gauge field''
       is described by $U(1)$ $2$-form gauge field $B^{(2)}$
       and charge-$N$ Higgs field;
       $U(1)$ should be broken to $\mathbb{Z}_N$.
 \item The topological charge $Q\pmod 1$ is formally given
       as~$-\frac{N}{8\pi^2}\int B^{(2)}\wedge B^{(2)}\in\frac{1}{N}\mathbb{Z}$.
       Note that this notation is not always correct
       because the de~Rham cohomology (or the $\wedge$ product)
       may miss the information of discrete gauge group.
       Therefore, we would replace $\wedge$ by the cohomological operations:
       $Q = -\frac{1}{N} \int_X \frac{1}{2} P_2(B^{(2)})\in\frac{1}{N}\mathbb{Z}$,
       where the Pontryagin square defined
       by $P_2(f) \equiv f\cup f + f\cup_1 \delta f\in H^{2q}(X,\mathbb{Z}_{2r})$
       with respect to $f\in H^q(X,\mathbb{Z}_r)$ possesses the graded commutativity
       for the cup product of simplicial cochains~\cite{Kapustin:2013qsa}.
       Note that we introduced the higher-cup product (e.g., $\cup_1$)
       because of the manifest $\mathbb{Z}_N^{[1]}$ gauge invariance
       at the cochain level.
 \item Finally, in the partition function with the $\theta$ term,
       we observe the violation of the $\theta$ periodicity
       \begin{align}
	\mathcal{Z}_{\theta+2\pi k}[B]
	= e^{2\pi i k Q}\mathcal{Z}_{\theta}[B],
	\qquad\text{with $Q\in\frac{1}{N}\mathbb{Z}$, $k\in\mathbb{Z}$}.
	\label{eq:mixed_anomaly}
       \end{align}
       This is called the mixed 't~Hooft anomaly
       between the $\mathbb{Z}_N^{[1]}$ symmetry and $\theta$ periodicity.
\end{enumerate}

\subsection{Topology of lattice gauge fields}
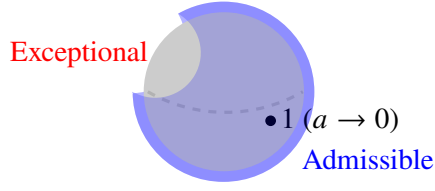
\begin{figure}[t]
 \begin{center}
 \begin{tikzpicture}
  \fill[blue!30] (-1.2,0) arc(-180:120:34pt) arc(30:-90:20pt);
  \fill[gray!40] (0,0) circle (30pt);
  \draw[dashed,very thick,gray!60] (-1,0) arc (-120:-60:58pt);
  \fill (1,0,1) circle (2pt) node[right] {$1$ ($a\to0$)};
  \fill[opacity=0.2,blue] (-1.2,0) arc(-180:120:34pt) arc(60:-120:17pt);
  \node[blue] at (1.9,-0.9) {Admissible};
  \node[red] at (-1.9,0.5) {Exceptional};
 \end{tikzpicture}
  \caption{Admissible or exceptional configuration space on $SU(2)\cong S^3$.
  By removing one point on~$SU(2)$ (in general a region),
  we have a well-defined configuration space depicted as the blue region,
  which is homeomorphic to a disk and called admissible.}
  \label{fig:admissibility}
 \end{center}
\end{figure}

Do we miss the topological structure on the lattice
because spacetime discretization breaks continuity?
The answer is no; L\"uscher~\cite{Luscher:1981zq}
constructed the $SU(N)$ principal bundle from lattice $SU(N)$ gauge fields
under the \emph{admissibility} condition such that
\begin{align}
 |1 - U_p| < \varepsilon \qquad\text{for $\exists\varepsilon>0$} .
\end{align}
This implies that $U_p$ should be sufficiently close
to the classical continuum limit~$\sim1$ because of its well-defined-ness;
see Fig.~\ref{fig:admissibility}.
Then, he proved the presence of topological sectors
and explicitly defined the topological charge on the lattice,
which takes an integral value.\footnote{%
It is known that the index can be computed on the lattice
thanks to the overlap Dirac operator~$D_{\mathrm{ov}}$, so that
\begin{align}
 \mathop{\mathrm{Index}}(D)
 = - \frac{1}{2}\Tr\gamma_5 D_{\mathrm{ov}} = n_{+} - n_{-} \in \mathbb{Z}.
\end{align}
Here $n_{\pm}$ is the number of positive/negative chiral modes.
The index theorem states that the index is identical
to the topological charge so constructed.}

Quite recently, in Refs.~\cite{Abe:2022nfq,Abe:2023ncy},
starting from the lattice action~\eqref{eq:wilson_action}
with the lattice $\mathbb{Z}_N$ $2$-form gauge field~$B_p$,
L\"uscher's construction was generalized;
we can find the fractional topological charge
and the mixed 't~Hooft anomaly~\eqref{eq:mixed_anomaly} on the lattice.
This generalized method provides a fully regularized framework
for studies of generalized symmetries and 't~Hooft anomaly matching.
We are ready to understand some studies more transparently and deeply.

To obtain an expression of~$Q$, we can use the Pontryagin square
which is defined \emph{on the hypercubic lattice}~\cite{Chen:2021ppt}.
On the other hand, after some calculations on an appropriate bundle structure,
\begin{align}
 Q[U_\ell, B_p] & = \sum_n q(n),&
 q(n) &= - \frac{1}{8N} \sum_{\mu,\nu,\rho,\sigma}
 \epsilon_{\mu\nu\rho\sigma}B_{\mu\nu}(n)B_{\rho\sigma}(n+\Hat\mu+\Hat\nu)
 + \Check{q}(n) ,
 \label{eq:topo_charge_lat}
\end{align}
where we have used $B_{\mu\nu}(n)$ instead of $B_p$ with $p=(n,\mu,\nu)$
and a unit vector of lattice $\Hat\mu$ in the $\mu$ direction.
We note that since $\sum_n\Check{q}(n)\in\mathbb{Z}$ the first term written
in terms of~$B_{\mu\nu}(n)$ gives rise to a fractional contribution.
For notational simplicity, let us introduce a product on the hypercubic lattice,
$\cup_{\mathrm{H}}$, as
$q(n)-\Check{q}(n)= -\frac{1}{2N}(B\cup_{\mathrm{H}} B)_n$ and so
$\sum_n\frac{1}{2}(B\cup_{\mathrm{H}} B)_n\in\mathbb{Z}$.\footnote{%
The definition of $\Check{q}(n)$ in Ref.~\cite{Abe:2023ubg} is quite complicated,
while we do not necessarily have to define it
in order to prove the fractionality~\cite{Abe:2023ncy}.
Actually, we can find
\begin{align}
 Q[U_\ell,B_p] \in - \frac{1}{8N} \epsilon_{\mu\nu\rho\sigma}
 z_{\mu\nu} z_{\rho\sigma} + \mathbb{Z},
\end{align}
in terms of the 't~Hooft flux~$z_{\mu\nu}$~\eqref{eq:twist_bc}.}
We can obtain the mixed 't~Hooft anomaly~\eqref{eq:mixed_anomaly}
within the lattice regularized framework.

\section{Instanton-sum modification and higher-group structure in lattice gauge theories}
\label{sec:higher-group}
Now, we add the term, $\sum_n i\chi(n) \left[q(n) - p c(n)\right]$,
in the Yang--Mills action~\eqref{eq:wilson_action}
with the $\theta$ term, $i\theta Q$.\footnote{%
For the $U(1)/\mathbb{Z}_q$ gauge theory,
the Witten effect for dyon plays an essential role
for the presence of not only higher-group but also higher-form symmetries.
Then, we should multiply $\theta$ by $q$
so that the $\theta$ term is given by $i q\theta Q$.
However it is not known under the admissibility condition
how to observe the monopole or dyon as dynamical degrees of freedom.
See recent works~\cite{Abe:2023uan,Aoki:2023lqp}.}
The equation of motion for the compact scalar~$\chi$,
which is a Lagrange multiplier field, implies that
the topological sectors are restricted to instanton numbers
as~$Q=p\sum_n c(n)\in p\mathbb{Z}$ for the $U(1)$ $4$-form field strength $c(n)$.
It is known that for any $p\in\mathbb{Z}$
this restriction provides a local and unitary quantum field theory.
Obviously, nontrivial configurations of~$B_p$ are forbidden
so that eq.~\eqref{eq:topo_charge_lat} takes a multiple of~$p$.
As we discuss from now,
gauging not only the $\mathbb{Z}_N^{[1]}$ symmetry (i.e., $B_p$)
but also the $\mathbb{Z}_p^{[3]}$ symmetry at the same time,
we can avoid this obstruction and observe the higher-group structure, that is,
$4$-group on the lattice~\cite{Kan:2023yhz,Abe:2023ubg}.

The fractional part of~$Q$ in the equation of motion is compensated
by a replacement as~$c-\frac{1}{Np}\Omega$,
\begin{align}
 q(n) - p c(n) + \frac{1}{N} \Omega(n) = 0.
\end{align}
We find that, assuming $\sum_n\Omega(n)\in\mathbb{Z}$,
all fractional contributions from~$B_p$ can be absorbed into~$\Omega(n)$.
There are alternative options as follows:
\begin{itemize}
 \item Strict structure:
       The minimal compensation is given
       by~$\Omega(n)=w(n)\equiv-N[q(n)-p c(n)]\in\mathbb{Z}$.
       The $\mathbb{Z}_N^{[1]}$ gauge transformation acts
       as eq.~\eqref{eq:1-form_gauge_transf},
       and the $\mathbb{Z}_{Np}^{[3]}$ gauge transformation acts as
       \begin{align}
	w(n)&\mapsto w(n)+(d\omega_{\mathrm{s}})_n\bmod Np, &
	c(n)&\mapsto c(n)+\frac{1}{Np}(d\omega_{\mathrm{s}})_n\bmod 1,
       \end{align}
       where $\omega_{\mathrm{s}}\in\mathbb{Z}$.
       Note that $w(n)$ is not transformed by $\mathbb{Z}_N^{[1]}$;
       the $3$-form symmetry is not $\mathbb{Z}_p$ but $\mathbb{Z}_{Np}$
       because the strict $4$-group mixes the $\mathbb{Z}_N^{[1]}$ symmetry
       into the \emph{physical} $\mathbb{Z}_p^{[3]}$ symmetry.
 \item Weak structure:
       The simpler equation of motion is realized by redefining $\Omega(n)$ as
       \begin{align}
	\Check{q}(n) - p c(n) + \Tilde\Omega(n) &= 0,&
	\Tilde\Omega(n) &\equiv \frac{1}{N}\Omega(n)
	- \frac{1}{2N} (B\cup_{\mathrm{H}} B)_n .
       \end{align}
       In the left hand side of the equation of motion,
       all terms contribute as integral values after the summation over~$n$.
       Then, we can see the explicit mixture of gauge transformations such that,
       thanks to eq.~\eqref{eq:1-form_gauge_transf} and
       \begin{align}
	\frac{1}{N}\Omega(n)
	&\mapsto \frac{1}{N}\Omega(n) + (d\omega_{\mathrm{w}})_n \bmod p, &
	c(n) &\mapsto c(n)+\frac{1}{p}(d\omega_{\mathrm{w}})_n \bmod 1 ,
       \end{align}
       where~$\omega_{\mathrm{w}}\in\mathbb{R}$,
       $\Tilde\Omega(n)$ transforms
       under the~$\mathbb{Z}_N^{[1]}$ and $3$-form gauge symmetries as
       \begin{align}
	\Tilde\Omega(n) \mapsto \Tilde\Omega(n)
	+ (d\omega_{\mathrm{w}})_n
	- \frac{1}{2N}\left[(B\cup_{\mathrm{H}}d\lambda)_n
	+(d\lambda\cup_{\mathrm{H}}B)_n
	+(d\lambda\cup_{\mathrm{H}}d\lambda)_n\right] ,
	\label{eq:3-form_gauge_transf}
       \end{align}
       where we have omitted the modulo operations for simplicity.
       By using the $\mathbb{Z}_N^{[1]}$
       and \emph{continuum} $3$-form gauge transformations,
       we suppose that $\Tilde\Omega(n)=\Tilde{w}(n)\in\mathbb{Z}$;
       that is, the genuine $3$-form symmetry is \emph{discrete} $\mathbb{Z}_p^{[3]}$
       as~$\omega_{\mathrm{w}}\in\mathbb{R}\to\mathbb{Z}$.
       The weak $4$-group is described in the local way by
       the mixture of the $\mathbb{Z}_N^{[1]}$
       and continuum $3$-form gauge symmetries,
       and the discrete $\mathbb{Z}_p^{[3]}$ gauge symmetry.
\end{itemize}

Finally, we make a remark on the case of the continuum theory.
As discussed in the previous section,
we have to use some more $U(1)$ gauge fields to describe this phenomenon.
Its explanation should be done in a careful way,
which may be quite wise but not transparent,
since there are some subtleties about the observation
of periodicity and the presence of higher-form symmetries.
On the other hand, on the lattice,
all we need to do is just taking a count of (fractional) numbers.

\subsection*{Acknowledgements}
The work of M.A. was supported by Kyushu University Innovator Fellowship Program in Quantum Science Area.
This work was partially supported by JSPS KAKENHI
Grant Numbers JP21J30003, JP22KJ2096 (O.M.) and JP22H01219 (N.K.).

\bibliographystyle{JHEP}
\bibliography{ref}

\providecommand{\href}[2]{#2}\begingroup\raggedright\begin{thebibliography}{10}

\bibitem{tHooft:1979rat}
G.~'t~Hooft, \emph{{Naturalness, chiral symmetry, and spontaneous chiral
  symmetry breaking}},
  \href{https://doi.org/10.1007/978-1-4684-7571-5_9}{\emph{NATO Sci. Ser. B}
  {\bfseries 59} (1980) 135}.

\bibitem{Gaiotto:2014kfa}
D.~Gaiotto, A.~Kapustin, N.~Seiberg and B.~Willett, \emph{{Generalized Global
  Symmetries}}, \href{https://doi.org/10.1007/JHEP02(2015)172}{\emph{JHEP}
  {\bfseries 02} (2015) 172} [\href{https://arxiv.org/abs/1412.5148}{{\ttfamily
  1412.5148}}].

\bibitem{Gaiotto:2017yup}
D.~Gaiotto, A.~Kapustin, Z.~Komargodski and N.~Seiberg, \emph{{Theta, Time
  Reversal, and Temperature}},
  \href{https://doi.org/10.1007/JHEP05(2017)091}{\emph{JHEP} {\bfseries 05}
  (2017) 091} [\href{https://arxiv.org/abs/1703.00501}{{\ttfamily
  1703.00501}}].

\bibitem{Sulejmanpasic:2020zfs}
T.~Sulejmanpasic, Y.~Tanizaki and M.~\"Unsal, \emph{{Universality between
  vector-like and chiral quiver gauge theories: Anomalies and domain walls}},
  \href{https://doi.org/10.1007/JHEP06(2020)173}{\emph{JHEP} {\bfseries 06}
  (2020) 173} [\href{https://arxiv.org/abs/2004.10328}{{\ttfamily
  2004.10328}}].

\bibitem{Morikawa:2022liz}
O.~Morikawa, H.~Wada and S.~Yamaguchi, \emph{{Phase structure of linear quiver
  gauge theories from anomaly matching}},
  \href{https://doi.org/10.1103/PhysRevD.107.045020}{\emph{Phys. Rev. D}
  {\bfseries 107} (2023) 045020}
  [\href{https://arxiv.org/abs/2211.12079}{{\ttfamily 2211.12079}}].

\bibitem{Abe:2022nfq}
M.~Abe, O.~Morikawa and H.~Suzuki, \emph{{Fractional topological charge in
  lattice Abelian gauge theory}},
  \href{https://doi.org/10.1093/ptep/ptad009}{\emph{PTEP} {\bfseries 2023}
  (2023) 023B03} [\href{https://arxiv.org/abs/2210.12967}{{\ttfamily
  2210.12967}}].

\bibitem{Abe:2023ncy}
M.~Abe, O.~Morikawa, S.~Onoda, H.~Suzuki and Y.~Tanizaki, \emph{{Topology of
  SU(N) lattice gauge theories coupled with \ensuremath{\mathbb{Z}}$_{N}$
  2-form gauge fields}},
  \href{https://doi.org/10.1007/JHEP08(2023)118}{\emph{JHEP} {\bfseries 08}
  (2023) 118} [\href{https://arxiv.org/abs/2303.10977}{{\ttfamily
  2303.10977}}].

\bibitem{Luscher:1981zq}
M.~L{\"u}scher, \emph{{Topology of Lattice Gauge Fields}},
  \href{https://doi.org/10.1007/BF02029132}{\emph{Commun. Math. Phys.}
  {\bfseries 85} (1982) 39}.

\bibitem{Kan:2023yhz}
N.~Kan, O.~Morikawa, Y.~Nagoya and H.~Wada, \emph{{Higher-group structure in
  lattice Abelian gauge theory under instanton-sum modification}},
  \href{https://doi.org/10.1140/epjc/s10052-023-11616-6}{\emph{Eur. Phys. J. C}
  {\bfseries 83} (2023) 481}
  [\href{https://arxiv.org/abs/2302.13466}{{\ttfamily 2302.13466}}].

\bibitem{Abe:2023ubg}
M.~Abe, O.~Morikawa and S.~Onoda, \emph{{Note on lattice description of
  generalized symmetries in $SU(N)/\mathbb{Z}_N$ gauge theories}},
  \href{https://doi.org/10.1103/PhysRevD.108.014506}{\emph{Phys. Rev. D}
  {\bfseries 108} (2023) 014506}
  [\href{https://arxiv.org/abs/2304.11813}{{\ttfamily 2304.11813}}].

\bibitem{Tanizaki:2019rbk}
Y.~Tanizaki and M.~\"Unsal, \emph{{Modified instanton sum in QCD and
  higher-groups}}, \href{https://doi.org/10.1007/JHEP03(2020)123}{\emph{JHEP}
  {\bfseries 03} (2020) 123}
  [\href{https://arxiv.org/abs/1912.01033}{{\ttfamily 1912.01033}}].

\bibitem{vanBaal:1982ag}
P.~van Baal, \emph{{Some Results for SU(N) Gauge Fields on the Hypertorus}},
  \href{https://doi.org/10.1007/BF01403503}{\emph{Commun. Math. Phys.}
  {\bfseries 85} (1982) 529}.

\bibitem{Kapustin:2013qsa}
A.~Kapustin and R.~Thorngren, \emph{{Topological Field Theory on a Lattice,
  Discrete Theta-Angles and Confinement}},
  \href{https://doi.org/10.4310/ATMP.2014.v18.n5.a4}{\emph{Adv. Theor. Math.
  Phys.} {\bfseries 18} (2014) 1233}
  [\href{https://arxiv.org/abs/1308.2926}{{\ttfamily 1308.2926}}].

\bibitem{Chen:2021ppt}
Y.-A.~Chen and S.~Tata, \emph{{Higher cup products on hypercubic lattices:
  application to lattice models of topological phases}},
  \href{https://arxiv.org/abs/2106.05274}{{\ttfamily 2106.05274}}.

\bibitem{Abe:2023uan}
M.~Abe, O.~Morikawa, S.~Onoda, H.~Suzuki and Y.~Tanizaki, \emph{{Magnetic
  operators in 2D compact scalar field theories on the lattice}},
  \href{https://doi.org/10.1093/ptep/ptad078}{\emph{PTEP} {\bfseries 2023}
  (2023) 073B01} [\href{https://arxiv.org/abs/2304.14815}{{\ttfamily
  2304.14815}}].

\bibitem{Aoki:2023lqp}
S.~Aoki, H.~Fukaya, N.~Kan, M.~Koshino and Y.~Matsuki, \emph{{Why magnetic
  monopole becomes dyon in topological insulators}}, {\emph{to appear in Phys.
  Rev. B} (2023) } [\href{https://arxiv.org/abs/2304.13954}{{\ttfamily
  2304.13954}}].

\end{thebibliography}\endgroup

\end{document}